\documentclass{article}

\usepackage{PRIMEarxiv}

\usepackage[utf8]{inputenc} 
\usepackage[T1]{fontenc}    
\usepackage{hyperref}       
\usepackage{url}            
\usepackage{booktabs}       
\usepackage{amsfonts}       
\usepackage{nicefrac}       
\usepackage{microtype}      
\usepackage{lipsum}
\usepackage{fancyhdr}       
\usepackage{graphicx}       
\graphicspath{{media/}}     
\usepackage{natbib} 

\usepackage{xcolor}
\usepackage{multirow}
\usepackage{caption}
\usepackage{subcaption}

\hypersetup{colorlinks,allcolors=black}

\title{Towards the design of user-centric strategy recommendation systems for collaborative Human-AI tasks
}

\author{
  Lakshita Dodeja\thanks{Authors contributed equally}, Pradyumna Tambwekar\footnotemark[1], Erin Hedlund-Botti, Matthew Gombolay \\
  School of Interactive Computing \\
  Georgia Institute of Technology \\
  Atlanta, GA\\
  \texttt{\{ldodeja3, pradyumnat, erin.botti\}@gatech.edu, matthew.gombolay@cc.gatech.edu} \\
   \And
}

\begin{document}
\maketitle

\begin{abstract}
Artificial Intelligence is being employed by humans to collaboratively solve complicated tasks for search and rescue, manufacturing, etc. Efficient teamwork can be achieved by understanding user preferences and recommending different strategies for solving the particular task to humans. Prior work has focused on personalization of recommendation systems for relatively well-understood tasks in the context of e-commerce or social networks. In this paper, we seek to understand the important factors to consider while designing user-centric strategy recommendation systems for decision-making. We conducted a human-subjects experiment (n=60) for measuring the preferences of users with different personality types towards different strategy recommendation systems. We conducted our experiment across four types of strategy recommendation modalities that have been established in prior work: (1) Single strategy recommendation, (2) Multiple similar recommendations, (3) Multiple diverse recommendations, (4) All possible strategies recommendations. While these strategy recommendation schemes have been explored independently in prior work, our study is novel in that we employ all of them simultaneously and in the context of strategy recommendations, to provide us an in-depth overview of the perception of different strategy recommendation systems. We found that certain personality traits, such as conscientiousness, notably impact the preference towards a particular type of system (p $<$ 0.01). 
Finally, we report an interesting relationship between usability, alignment and perceived intelligence wherein greater perceived alignment of recommendations with one's own preferences leads to higher perceived intelligence (p $<$ 0.01) and higher usability (p $<$ 0.01). 
\end{abstract}


\section{Introduction}

The increasing capabilities of AI-systems has led to their widespread use in many fields. However, for safety-critical domains, such as search and rescue~\citep{murphy2004human,heintzman2021anticipatory}, aviation~\citep{li2021human} and healthcare~\citep{magrabi2019artificial}, a partnership between humans and AI is preferred over complete reliance on AI-systems.
To engender effective collaboration, humans need to be able to specify their intent with regards to how an AI system should perform the collaborative task, such that both human and AI-partners work towards the same goals~\citep{gombolay2017computational,tabrez2020survey, nikolaidis2012human,johnson2021interdependence}. 
On the other hand, humans may not always be able to communicate their intention as executable instructions, due to a lack of technical expertise required to structure their intent into the appropriate format.
In such situations, it may be helpful for the AI-system to query for a user's intrinsic preferences and present a recommendation for an executable strategy that the human can accept, reject, or modify. 
This process would enable humans who are non-experts to coordinate on the strategies for the task they are trying to collaboratively accomplish. Motivated from prior work, we formulate the strategies in terms of goals and constraints required to achieve the task \citep{https://doi.org/10.48550/arxiv.2208.08374}.
We use the term ``strategy recommendation systems" to define these systems in our work.

To design these strategy recommendation systems, we need to consider and incorporate the functional and dispositional requirements of end-users. Prior studies on recommendation systems and human-AI collaboration elucidate the need for the user-centric design of such systems to engender effective collaboration.
Users have been shown to be more resistant to utilizing generic AI systems since they would not be able to cater to the unique characteristics and demands of the users \citep{longoni2019resistance}. 
Personalizing recommendations is one such method of enabling AI systems to curate suggestions that are specific to a user \citep{huang2022effects}. 
Personalization has already proven to be effective in increasing consumer satisfaction \citep{xiao2019beyond} and business revenues \citep{behera2020personalized}.
Previously, content personalization for Recommendation Systems based on user preferences has been done through collaborative filtering, content-based filtering and hybrid approaches \citep{kumar2018recommendation, he2017neural, schafer2007collaborative, lops2011content}.

Developing user-centric recommendation systems not only involves personalizing the content of recommendations but also the way content is presented to the users through a personalized user interface or interaction process.
Despite these algorithmic advances in content personalization, humans have nuanced requirements that can impact their interactions with the recommendation system.
Sometimes humans find it useful being recommended items relevant to their needs and preferences, such as during e-learning \citep{tarus2018knowledge} and for e-commerce \citep{linden2003amazon}. 
Other times, humans seemed to be more satisfied upon receiving a diverse set of options to choose from \citep{kunaver2017diversity}.
Personality traits have also shown to impact the inclination towards a particular structure of items within a recommendation \citep{nguyen2018user}. For instance, Nguyen et al. showed that users who are more introverted preferred diverse recommendations over similar recommendations. Also, users who were less open preferred recommendations that were more in line with their previously consumed data. 
However, these findings might not be consistent in the context of recommendation systems for Human-AI collaboration.
Therefore, it is important to study how the structure of the items in recommendation impact the perception of the strategy recommendation system.

In this paper, we propose a novel humans-subjects experiment to understand factors that would be relevant for designing effective strategy recommendation systems for human-AI partnerships in safety-critical domains. Our work seeks to study how we can best personalize the collection of strategy recommendations for humans. 
Unlike prior work studying personalization within recommendation systems, which studies the relevancy of content within a recommendation itself, we seek to understand how to best select the assortment of strategy recommendation(s) to suit a user's personality and disposition. We want to determine whether the nature and structure of strategy recommendations influence a user's perception of the system.
For our study, we develop a validated metric for converting a user's gameplay preferences into actionable strategies. 
We employ this metric to present recommendations to a user based on their preferences. In our experiment, we study four baselines motivated from prior work \citep{tan2008learning,szpektor2013relevance,bollen2010understanding} but through a between-subjects design and in the context for strategy recommendation systems; (1) Single strategy recommendation which most closely align with a user's preferences, (2) Multiple recommendations which are similar to a user's preferences, (3) Multiple recommendations which include strategies both similarly and oppositely aligned with their preferences, (4) All possible strategies.
The benefit of the between-subjects design is that it allows us to compare and contrast all the strategy recommendation systems through the same metrics.
We study how personality type and predisposition towards preferring certain types of recommendations affects a user's alignment, preference and perceived intelligence of any given strategy recommendation type. Our overall contributions are as follows:
\begin{enumerate}
    \item We design a novel user study to understand user-preferences with respect to strategy recommendations for challenging tasks. 
    \item We develop and validate a metric to convert a user's preferred gameplay style into an actionable strategy using just three questions (p $<0.05$). 
    \item We evaluate the user preference in the form of usability and perceived intelligence of the system, workload for the task, and the alignment of the recommendations with user's strategy (p $<0.01$).
    \item We study how perceived alignment of the recommendation list impacts the general perception of a recommendation type. We found that perceived alignment of the recommended options with user preference significantly impacts the usability (p $<0.01$) and perceived intelligence (p $<0.01$) of the system. 
\end{enumerate}
Our first-of-its-kind human-subjects experiment provides a deeper understanding of factors that impact human preferences while interacting with a strategy recommendation system for decision making. These insights provide a foundation for designing strategy recommendation systems that can adapt to user preferences.

\section{Related Work}
In this section we will first cover the existing methods for personalization within a Recommendation System. 
We will then discuss different works studying the impact of the presentation of items within a recommendation systems on users and perception of recommendation systems.
\subsection{Recommendation System and Personalization}
Recommendation Systems can be personalized or unpersonalized based on the task requirement. Personalization in recommendation has been mostly achieved through Collaborative Filtering~\citep{sarwar2001item, kluver2018rating} and Content-based Filtering~\citep{van2000using, vanetti2010content}.  
Collaborative Filtering finds  another user with similar preferences and performs recommendations based on the neighboring user's preferences while content-based filtering performs recommendations based on past preferences of the user \citep{koren2022advances}. 
Collaborative Filtering can be done through the neighborhood approach or latent factor models. Neighborhood approaches work by either finding similar users or similar items whereas latent factor models try to find similar characteristics by factorizing users and items in the same embedding space.
Content based filtering involves either discovering new data sources like linked open data, user generated content or new algorithmic approaches like meta-path based approaches, encoding metadata and deep learning~\citep{lops2019trends}. Explicit item rating, adaptive dialogues and forms, and comparison based techniques have been used to elicit user preferences. 
Interactive techniques like visualization, explanations, and user feedback can further help in shaping the recommendation interface based on user preference~\citep{jugovac2017interacting}. 
The personalization in Recommendation Systems can be done through User Interfaces, Content or the Interaction Process \citep{zanker2019measuring}. Rather than studying the algorithmic accuracy of the content presented to a user, in our study we seek to understand how the user interface can be personalized for strategy recommendation systems.
 


\subsection{Presentation of items within Recommendations}

Recommendation systems generally present options which are similar to user preferences~\citep{tan2008learning,chen2006recommendation, linden2003amazon} but in some cases it can be beneficial to have  options which are dissimilar to user preferences. ``Freshness and Diversity" in recommendations has proven to be helpful in improving the performance of the personalized question recommendation system~\citep{szpektor2013relevance}. A recent study on music recommendation systems also pointed towards the need for algorithms that are diversity aware, in other words, algorithms that are able to recommend relevant as well as diverse music options to users~\cite{anderson2020algorithmic}.

Prior work has also found that the lack of transparency and unpredictability in algorithms can sometimes lead to users feeling helpless while working with automated systems~\citep{jhaver2018algorithmic}, while expertise of the system was proven to be helpful towards a human's appreciation of the system  \citep{hou2021expert}. Thus, showing all kinds of strategies might help users understand the recommendation system better and in turn reduce their anxiety while using the system. Conversely, it could further lead to ``choice overload"~\citep{bollen2010understanding}. Choice overload happens when the recommendation system provides a large set of good options for users to choose from. It has been shown as the number of options increases the satisfaction of the users increases marginally and then starts to decrease \citep{reutskaja2009satisfaction}. 

In our study, we consider these phenomenon in designing four types of recommendations to present to users, i.e. single aligned recommendation, similar aligned recommendations, three diverse recommendations comprised of both similarly and oppositely aligned strategies, and all possible recommendations. However, unlike prior work, we compared all the types of recommendation systems simultaneously and in the context of strategy recommendations. 
This lets us compare and contrast different strategy recommendation systems and their perception across users.
We also analyzed the correlation between personality traits of the user and preference towards a particular recommendation system.
We hope to study whether prior findings on diversity, anxiety, and choice overload are reflected in user-interactions with strategy recommendations. 


\subsection{Studying human perception of Recommendation Systems }

Recommendations Systems should not only be able to accurately estimate the similarity and dissimilarity between two items but also be able to generate user-centric suggestions. The concept of Human-Recommender Interaction advocates redesigning the recommendation system from the end user's perspective to better meet their needs~\citep{mcnee2006making}. Since then, many user studies have been conducted to understand how users perceive different recommendation systems and how recommendation systems can impact users' decision making. Studies have been conducted to determine how recommendation list should be displayed, whether the list should be personalized and for what scenarios  \citep{tam2003web} or whether it should be diverse~\citep{  willemsen2016understanding,ziegler2005improving}. 
Prior work has also conducted experiments to analyze the impact of recommendations by a physical robot and on screen agents on human decision making  \citep{shinozawa2005differences}. 
They showed that a three dimensional figure is not always beneficial over a two dimensional figure while communicating recommendations as it also depends on the interaction environment being used.

Various guidelines have also been established to conduct these user studies that help researchers in determining the system aspects to be studied, dependent variables to be used, etc.~\citep{knijnenburg2015evaluating}. 
Human-subjects experiments also serve as an important method to evaluate the efficacy of a recommendation systems~\citep{shani2011evaluating}. We can test new systems by recommending items generated by the systems to humans and checking if the humans perceive them better than recommending them random items or items from a different algorithm.
We seek to leverage the insights from these prior experiments, to design our human-subjects experiment and analyze the perception and usability of strategy recommendation systems. In our work, we perform a first-of-its-kind human-subjects experiment to evaluate and understand a user-centric strategy recommendation system.



\section{Experimental Design} 
We designed a novel human-subjects study to understand user-preferences with respect to recommendations of game-play strategies. 
Through our study, we seek to understand what factors influence how humans like to be provided strategy recommendations, such as their predisposed proclivity towards a specific type of recommendation or their personality.
Ours is the first experiment to study user perceptions of recommendation systems outside of well-defined tasks such as e-commerce or social media recommendations. 
We seek to provide design insights for recommendation systems built for human-AI collaboration, in order to enable AI-systems to suggest strategy recommendations to reduce the cognitive load of the human-collaborator. 

Our strategy recommendations are defined in terms of goals and constraints. 
Goals are defined as the set of desirable states that you want to achieve and constraints are the conditions imposed while obtaining those states (see Figure~\ref{fig:actual_reverse} for an example strategy).
We chose this scaffolding of goals and constraints here because goals and constraints are effective mechanisms for programming an AI-agent's behavior. They can be easily plugged into both learning and planning-based methods to specify an agent's task.
For a strategy, each of the six goals has a value between unfavorable, neutral, and favorable which dictates the importance of the goal towards the overall strategy. 
Constraints are represented as individual statements, such as ``I need 4 troops to effectively defend a country,'' or ``I must protect the borders of Asgard.'' 

\subsection{Environment}
We utilized the board game Risk, for our experiment. 
Risk was an ideal environment for this study, due to the various contrasting strategies that can be employed towards winning a game of Risk. 
This environment is also complex enough to necessitate a strategy recommendation system, as without any recommendations the player would need to play several rounds of the game in order to curate their own strategy.
Playing Risk involves completing various resource allocation, scheduling, and planning tasks which are key parts of many real-world tasks.
Furthermore, Risk is a stochastic environment, which is congruent with real-world scenarios, like financial trading, disaster response, search and rescue, robot manipulation, etc.
Unlike other scheduling or resource-allocation games, such as Starcraft or Age of Empires, humans can more intuitively develop and interpret strategies for Risk. Risk has a significantly simpler rule-set which enables humans to develop and understand strategies without large amounts of domain expertise.
These properties make Risk a suitable environment to understand potential trends for recommending AI-strategies for real-world tasks. 

\begin{figure}[t]
  \centering
  \includegraphics[width=0.7\linewidth]{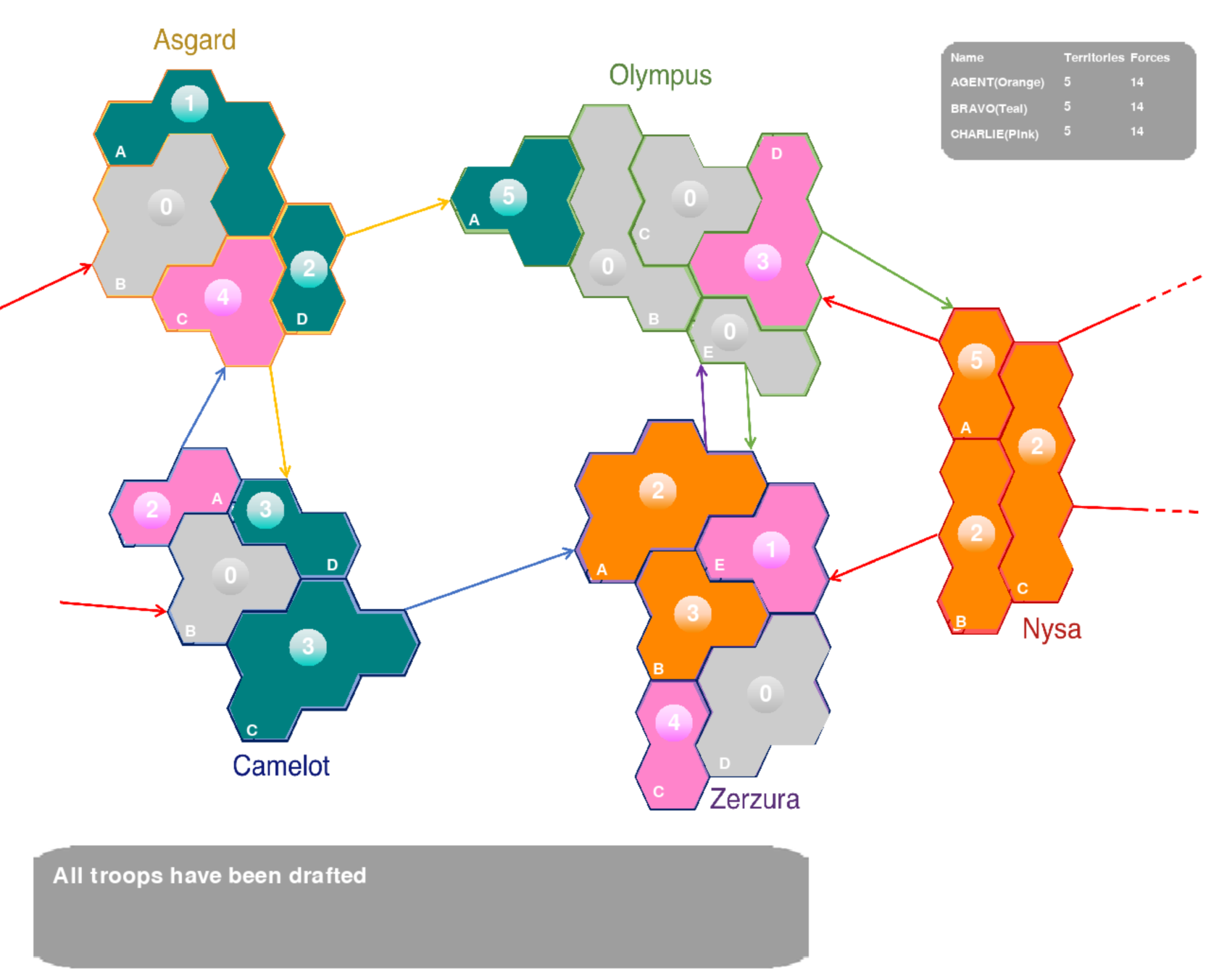}
  \caption{This figure shows the Risk Simulator used for our study. Simulation for the recommended strategy was executed by the orange player (Agent) which was playing against teal (Bravo) and pink (Charlie) players. We also included a legend so that participants could track the forces and territories of all players. Each action is annotated with a text-description in the text box at the bottom of the screen.}
  \label{fig:risk_simulator}
\end{figure}

The version of Risk we employ is a turn-based game with three players that is comprised of four primary phases:
\begin{enumerate}
    \item \textbf{Draft}: Pick your initial set of territories on the game board and deploy your initially allocated troops.
    \item \textbf{Reinforce}: Deploy additional troops to your existing territories.
    \item \textbf{Attack}: Conduct battles between territories you control with opposing territories.
    \item \textbf{Maneuver}: Move troops between two territories you control. 
\end{enumerate}
The draft phase is conducted only once per player at the start of the game. After the initial drafting phase of all players, each player's turn is comprised of sequentially completing the \textit{Reinforce}, \textit{Attack}, and \textit{Maneuver} phases. 
A player wins when they have conquered all territories on the map. For our experiment, we adapt a Risk gameplay simulator from prior work~\cite{https://doi.org/10.48550/arxiv.2208.08374}. 
We created gameplay agents within this simulator that could follow the strategies we developed for this study, as a means of providing participants with a practical demonstration of what deploying any given strategy looks like. Our Risk simulator is shown in Figure~\ref{fig:risk_simulator}.


\subsection{Research Questions}
Through our experiment, we sought to answer the following research questions: 
\begin{enumerate}
    \item[RQ1] - How can we automatically calibrate an actionable strategy which aligns with a participant's gameplay preferences?
    \item[RQ2] - Does our study condition, i.e. the type of strategy being recommended to users, have an impact on the dependent variables for the recommendation system?
    \item[RQ3] - How do personality factors and demographic information influence the dependent variables in this study? 
    \item [RQ4] - How does perceived alignment of the recommendation list impact the general perception of a recommendation type? 
\end{enumerate}
Our experiment is divided into two phases. In Phase 1, we validate our mechanism to recommend a strategy aligned with a participant's preferences (\textit{RQ1}). In Phase 2, we study the perception of varying mechanisms of recommending strategies, and identify the factors that impact these perceptions (\textit{RQ2-4}). 


\section{Learning Preferred Strategy - Phase 1}
\label{sec:prefQ}
The first step in our experiment is to identify a mechanism to encode a participants innate gameplay preferences into an aligned actionable strategy (\textit{RQ1}).
To accomplish this goal, we require a meaningful set of questions to ascertain a participant's gameplay preferences. 
First, we developed eight different strategies that represent a diverse range of gameplay strategies for Risk.
Next, we developed a questionnaire, wherein the participant answered three questions regarding what action they would take in fictional scenario in Risk. 
This questionnaire was structured as a decision tree of depth three, such that the question a participant receives is based on their answer to the previous question (see Figure~\ref{fig:rec_example}).
Through their answers to the three questions presented, we ascertain which out of the eight strategies (i.e., $2 ^ 3 = 8$ options) most aligns with how the participant would play the game. 
Prior to deploying our questionnaire in our study, we first developed a study to validate whether our method can accurately calibrate a participant's preferred Risk strategy.
This section covers the analysis of pilot studies which informed the final structure of the validation study as well as the final procedure employed to validate our questionnaire. 

\subsection{Pilot Studies}
\begin{figure}[t]
  \centering
  \includegraphics[width=\linewidth]{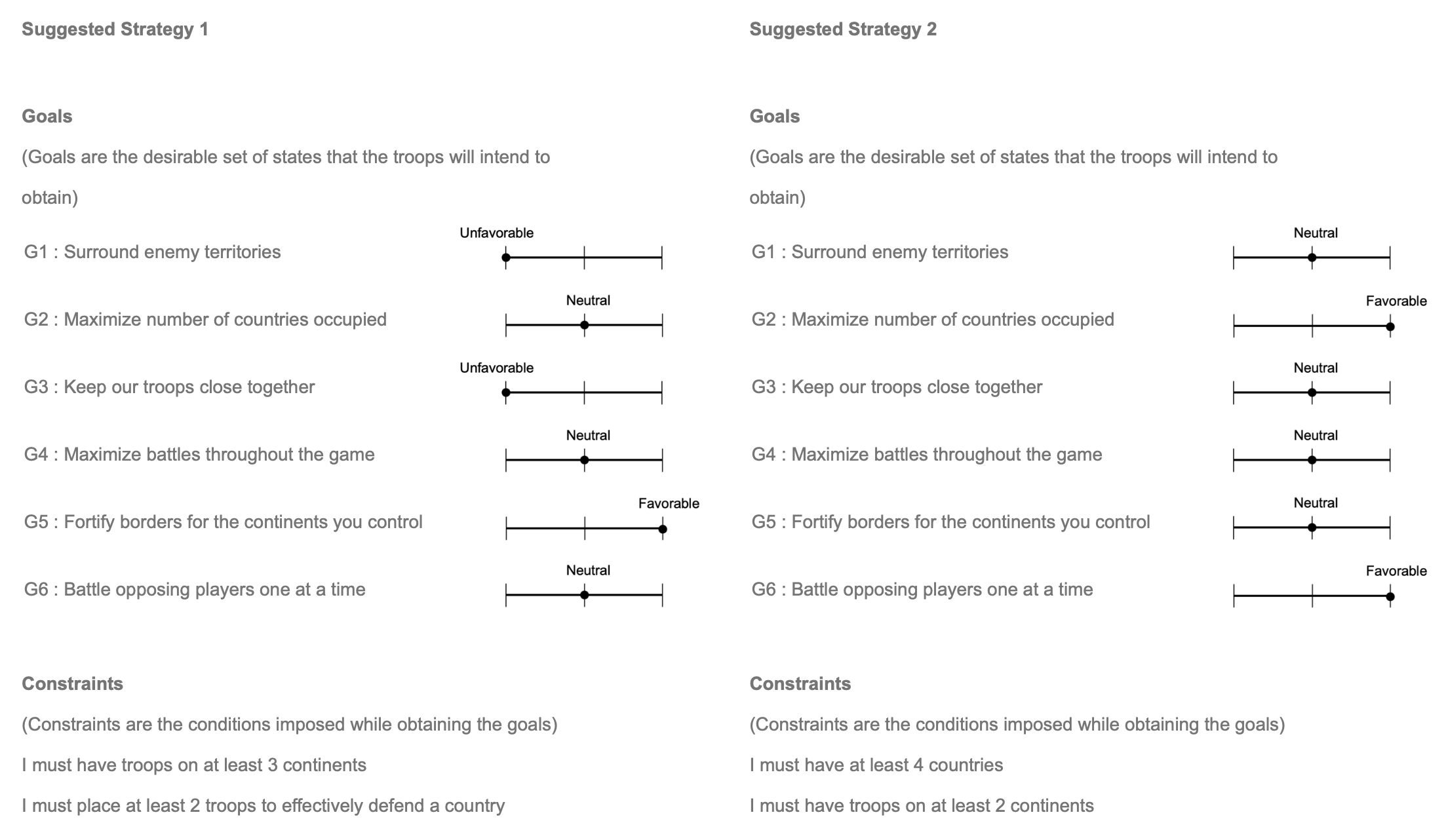}
  \caption{This figure depicts how the actual and reverse strategies were presented to participants during the calibration study. The first half of participants were shown the actual strategy as ``Strategy 1'' and the second half of participants were shown the reverse strategy as ``Strategy 1.'' }
  \label{fig:actual_reverse}
\end{figure}

We iterated over our study design for the calibration phase several times during the piloting phase to identify and fill the gaps in our study. In the first iteration, we only displayed the strategy produced by our decision tree (actual strategy (AS)) to the participants and asked them to fill out the alignment questionnaire (see Table~\ref{tab:alignment}) but soon realised this might lead to confirmation bias in our study (n=12, mean=5.4375, SD=1.28). In the second iteration, we incorporated both the actual as well as reverse of that strategy (RS) in our study. The participants were first recommended one of the strategies and asked to fill the alignment questionnaire for that strategy then the same process was repeated for the next strategy. Since participants were not able to compare the two strategies they ended up giving high alignment scores to both the strategies (n=3, AS: mean=6.33, SD=0.62, RS: mean=5.42, SD=1.19). Therefore, in the next phase we displayed both the strategies side by side (Figure~\ref{fig:actual_reverse}). Participants were now able to compare and contrast the two strategies. People who preferred the reverse strategy were not able to coherently explained their choices (n=5, AS: mean=4.45, SD=1.94, RS: mean=5.5, SD=1.20). So, in the final phase of the study we asked the participants to explain their strategy behind their choices in the preference questionnaire before displaying the recommended strategies to them. Also, we recruited participants who already had some experience with playing strategy games like RISK, CATAN, etc. so they could validate our preference question better. If the participants still preferred the reverse strategy over the actual strategy we asked them some semi-structured interview questions (Appendix C). 

\begin{table}
  \centering
  \caption{This table depicts the likert statements employed in our alignment questionnaire. Each statement has a response format with seven items, ranging from strongly disagree to strongly agree.}
  \vspace{15px}
  \label{tab:alignment}
  \begin{tabular}{cl} 
    \hline
    \toprule 
    &Likert Scale\\  \hline
    \midrule
     1& The suggestions made to me were \textbf{aligned} with my strategy\\ 
     2& The suggestions \textbf{failed} to capture my strategy\\ 
     3& The suggestions \textbf{paid attention} to my strategy\\ 
     4& The suggestions \textbf{ignored} my preferred strategy\\ \hline
  \bottomrule
\end{tabular}
\end{table}

\subsection{Validating Calibration Questionnaire}
We began our study with a tutorial for the rules of Risk to familiarize participants with the rules of the environment, and the simplifications we made to the game. 
Next, participants were allowed to play a Risk simulator to familiarize them with how the gameplay works. 
The study administrator would walk participants through each phase of the game within the simulator, and then participants would be allowed to play the game until they are confident about their understanding of the game. 
As in each of our pilot studies, our questionnaire was a binary tree, consisting of multiple-choice questions based on fictional scenarios in Risk. 
Each question had two options and based on their answer they would be provided a different question.
Every question seeks to answer specific questions regarding their gameplay preferences, such as ``Are they an aggressive player?'', ``Do they like to keep their troops close together?'', ``Do they prefer to maximize battles or countries controlled?'', etc.   
We have included each of the seven questions in our tree, as well as the structure of the tree itself within the Appendix. 
After the participant answers three questions, our method gives us a unique strategy profile that aligns with how the participant would play Risk. 
We developed a total of eight gameplay profiles, each of which had an associated heuristic-based gameplay agent in order to simulate how each profile would function in practice. 

Next, we present each participant with two options (1) The ``actual'' strategy, which corresponds to the strategy that best aligns with their answers to the questionnaire, (2) The ``reverse'' strategy which oppositely aligns with their answers to the questionnaire, i.e. is on the opposite side of the decision tree. Both strategies are presented side-by-side to participants.
Participants alternatively received the actual strategy or the reverse strategy as the Suggested Strategy 1 first (see Figure~\ref{fig:actual_reverse}).
We then showed the participant a simulation of an AI Risk player which employs the given strategy and asked them to answer an alignment questionnaire, comprised of four questions (see Table~\ref{tab:alignment}) to gauge how well the strategy aligned with their preferences. 
This survey was evaluated on a 7-point, 4-item, Likert scale from Strongly Disagree to Strongly Agree. 
Finally, the participants were asked subjective questions regarding what they liked and disliked about the suggested strategy.
This process was conducted sequentially for both strategies, i.e. simulation, alignment, subjective for strategy 1, then simulation, alignment subjective for strategy 2. We discarded one data point where the participant acknowledged that the actual strategy was a better representation of his choices in the scenarios but rated the reverse strategy higher.

\subsection{Results}
In this section, we share the statistical results from our calibration study to validate \textit{RQ1}.
We conducted our Calibration Study with 16 participants and had to discard data for one participant. Out of the final 15 participants, 8 of them saw the actual strategy as Strategy 1 (left side in Figure~\ref{fig:actual_reverse}) while 7 participants saw the reverse strategy as Strategy 1. On average the participants rated the alignment for the actual strategy (Mean = $5.18$, SD = $1.28$) higher than the reverse strategy (Mean = $3.93$, SD = $1.78$). The data failed the Shapiro-Wilk Test for normality thus we used the Wilcoxon signed-rank test. A Wilcoxon signed-rank test confirmed that this difference was statiscally significant with Z $= -3.22$ and p $<0.05$. The Cronbach's alpha for the alignment questionnaire was $\alpha = 0.94 $. 
The statistical significance of the Wilcoxon signed-rank test confirms that the strategies recommended by our questionnaire are accurately aligned with a participant's preferences. 
This result validates our assumption that this questionnaire encodes preferences into actionable Risk strategies for Phase 2 of our experiment.

\section{Phase 2 - Strategy Recommendation Preference Study}
After validating our preference questionnaire, we move on to our second experiment wherein we validate \textit{RQ2-4}. We conducted our second study as an in-person experiment, with a $1 \times 4$ between-subjects design wherein the study condition is the method of recommending a strategy to a user. 
We first calibrate a participants gameplay preferences through the preference questionnaire from Section~\ref{sec:prefQ}.
Following this calibration, we recommend strategies to users in one of the following four formats:
\begin{enumerate}
    \item \textbf{Single} - Participants are recommended a single strategy that best aligns with their preferences regarding how they would play Risk.
    \item \textbf{Similar} - Participants are recommended three strategies that are all similar and aligned with their gameplay preferences.
    \item \textbf{Diverse} - Participants are recommended three strategies which include strategies which are similar and opposing to their gameplay preferences. 
    \item \textbf{All} - Participants are recommended all possible strategies for playing Risk. In our study we have eight total strategies. 
\end{enumerate}

\subsection{Procedure}
We started the study in a similar manner as the calibration study where we provided the participants with a tutorial of RISK and the RISK simulator to explore and play the game. 
Prior to recommending any strategies, participants were asked to fill out the Mini-IPIP personality questionnaire, to measure their personality type~\citep{donnellan2006mini}. 

\begin{figure}[t]
  \centering
  \includegraphics[width=0.9\linewidth]{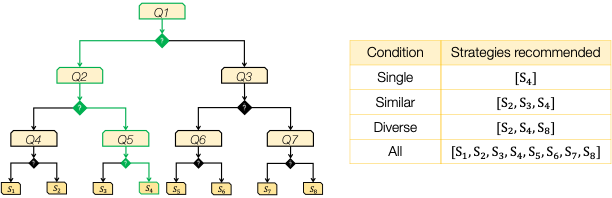}
  \caption{A depiction of how we recommend strategies based on the study condition assigned to the participant. In this illustration, based on the participant's answers to each question, their ideal strategy is $S_4$, as shown by the path highlighted in green. For the ``single'' condition, we will recommend only $S_4$. For the ``similar'' condition, the participant is recommended 3 strategies, i.e. the sibling strategy, $S_3$, and one of its ``cousin'' strategies, $S_2$. In the ``diverse'' condition, the participant is also recommended 3 strategies, however instead of the sibling strategy, the participant is recommended a strategy on the other side of the tree, i.e. $S_8$. Finally, with respect to the ``all'' condition, participants are shown all eight strategies. The full calibration questionnaire can be found in Appendix B.1.}
  \label{fig:rec_example}
\end{figure}

After completing the tutorial and pre-survey segments, participants were allowed to begin the study. The study started with participants filling out the calibrated preference questionnaire described above to curate their ideal strategy.
Before providing our strategy recommendations, participants are asked to explain verbally their reasoning for their choice in each scenario. 
This serves as a mechanism for participants to re-calibrate their holistic strategy based on their choices, and serves as a mechanism to filter out participants who made a mistake or no longer agree with their original choices. 
Next, we present the one, three, or eight strategy recommendations to the participant based the study condition they have been assigned. Figure~\ref{fig:rec_example} provides an illustration of how strategies are recommended based on the participant's answers to each question in the strategy questionnaire. 
The strategies are presented in order of relevance, i.e. most related to their gameplay preferences to least related to their preferences, but participants were not informed of the ordering. 
In addition to the goals and constraints within each strategy, we also provided an image how a player following the given strategy would have drafted troops in the drafting phase (Appendix Fig 2). 
Finally, for each recommended strategy, the participant had the option of viewing a simulation of the first two turns of gameplay with an agent that utilizes the strategy. The simulation helped the participants visualise the particular strategy if they had difficulty in parsing goals and constraints of the strategy.
We restricted our simulation to two turns as we did not want the participant to be biased by whether the gameplay agent wins. 
Our aim was to encourage participants to evaluate their recommendations based on their alignment towards how the participant would play the game rather than how successful a strategy is.
If participants see a strategy fail, which is possible due to the stochastic nature of Risk, they may be less likely to admit that the the strategy aligned with how they would play the game. 

After participants confirm that they have understood each of the recommended strategies, the participant is asked to answer four post-experiment questionnaires regarding their experience working with our recommendation interface. 
First, they fill out an alignment questionnaire, which utilizes the same questionnaire employed in our calibration study (see Table~\ref{tab:alignment}). 
Next, participants fill out the system usability survey from prior work~\citep{brooke1996sus}, to estimate how usable participants felt the recommendation interface was. 
After evaluating usability, the participants filled out the NASA Task Load Index (TLX)~\citep{hart1988development} survey used to measure the workload for the task. Next, the participants filled out the Godspeed perceived intelligence questionnaire~\citep{bartneck2009measurement} to assess the intelligence of the recommendation process. 
Finally, the last questionnaire was a novel preference questionnaire, which sought to understand a participants general preferences with respect to receiving \textit{single}, \textit{similar}, \textit{diverse}, or \textit{all} recommendations (Appendix B.2). 
Each category had four associated questions, with a total of 16 Likert items. 
This survey was evaluated on a 7-point Likert response format from Strongly Disagree to Strongly Agree. 
We also included two subjective questions, i.e. (1) What did you like about the Recommendation Interface? and (2) What did you dislike about the Recommendation Interface? 
Through these questions, we hoped to obtain qualitative insights regarding participants' opinions and experience with respect to each type of recommendation. These questions were asked before our novel preference questionnaire to eliminate any bias. 

\subsection{Results}
In this section, we report the statistical tests conducted in both the calibration study and the main study to answer the research questions setup in this paper. 
We ran our experiment with a total of 60 participants. As a post study survey, we collected additional demographic information from the participants. Out of the 58 participants that reported, 21 were female, 36 were male and 1 was Non-Binary. The average age of participants was 23.5 (s.d. - 2.95) and the average expertise in playing strategy games on a scale of 1-5 was 3.08 (s.d. - 0.98).
The estimated time for our study was 45 minutes and participants were paid $\$$15 to participate in our study.
We discuss the statistical tests conducted to study each research question and analyze our findings with respect to user-preferences for strategy recommendations.

To answer our research questions regarding user preferences of strategy recommendations (\textit{RQ2 - 4}), we performed a multivariate regression analysis for each dependent variable. 
For all linear regression models, we tested for the assumptions of normality of residuals and homoscedasticity. 
We performed Levene's test measure homoscedasticity, and all models were found to be homoscedastic. The summary of all the models and assumption tests can be found in the Appendix.
We performed Shapiro-Wilk's test to test for normality of the residuals of each model, however we found that some models had residuals that were not normally distributed. 
In prior work, it has been shown that an F-test is robust to non-normality~\citep{cochran1947some, blanca2017non, hack1958empirical, glass1972consequences}. Therefore, we choose to proceed with a linear regression analysis. 
To find the appropriate model for each  measure, we applied AICc as our occam's razor. Owing to our comparatively small sample size (n=60), we used AICc which adds a correction term to the standard AIC to avoid overfitting.
We performed a one-way ANOVA to measure significance of each measure on the dependent variable. 
We further conducted a TukeyHSD post-hoc test to identify pairwise significance between values for independent variables which were structured as factors. 


\begin{figure}[t]
     \centering
     \begin{subfigure}[b]{0.48\textwidth}
         \centering
         \includegraphics[width=\textwidth]{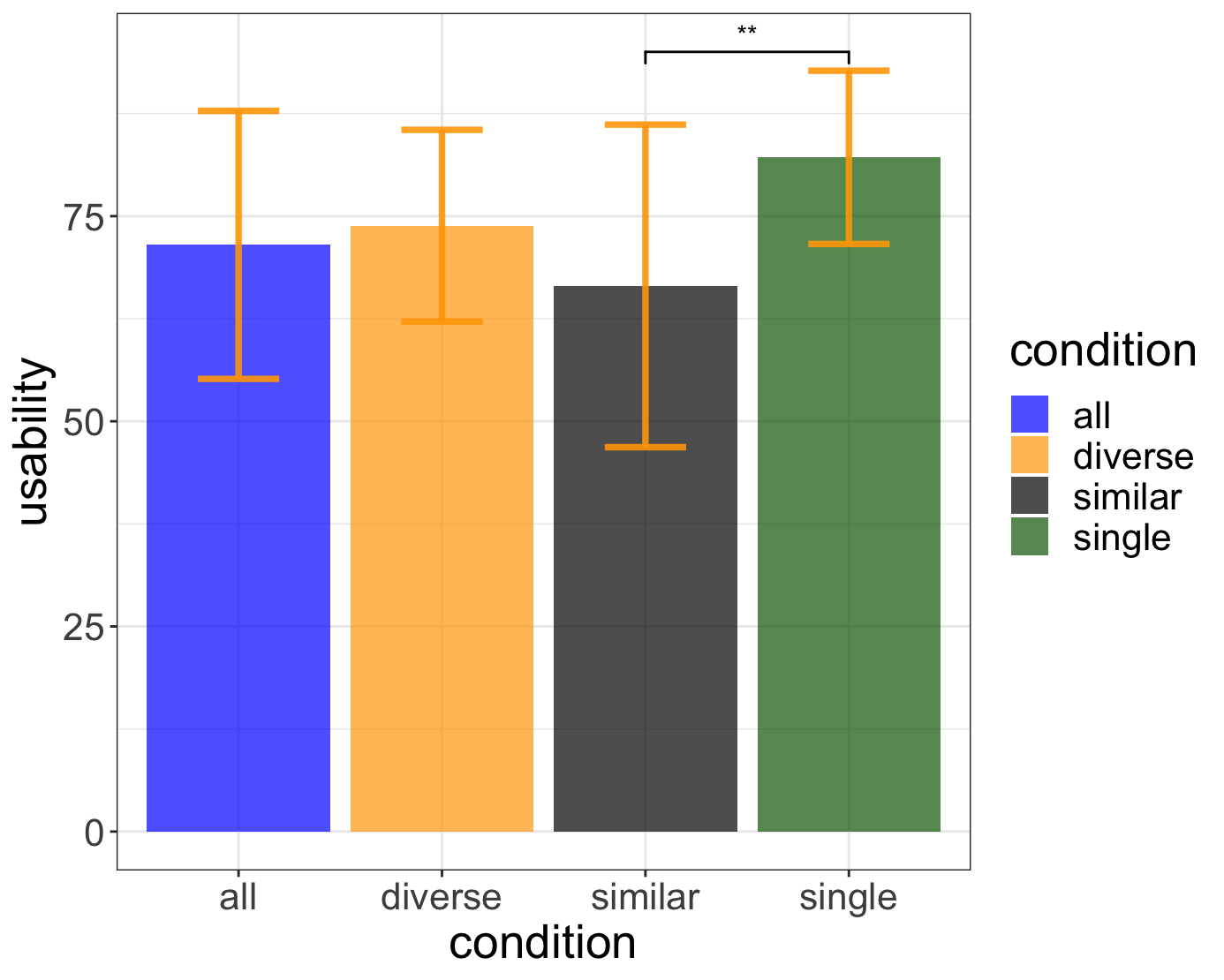}
         \caption{}
         \label{fig:usability}
     \end{subfigure}
     \hfill
     \begin{subfigure}[b]{0.48\textwidth}
         \centering
         \includegraphics[width=\textwidth]{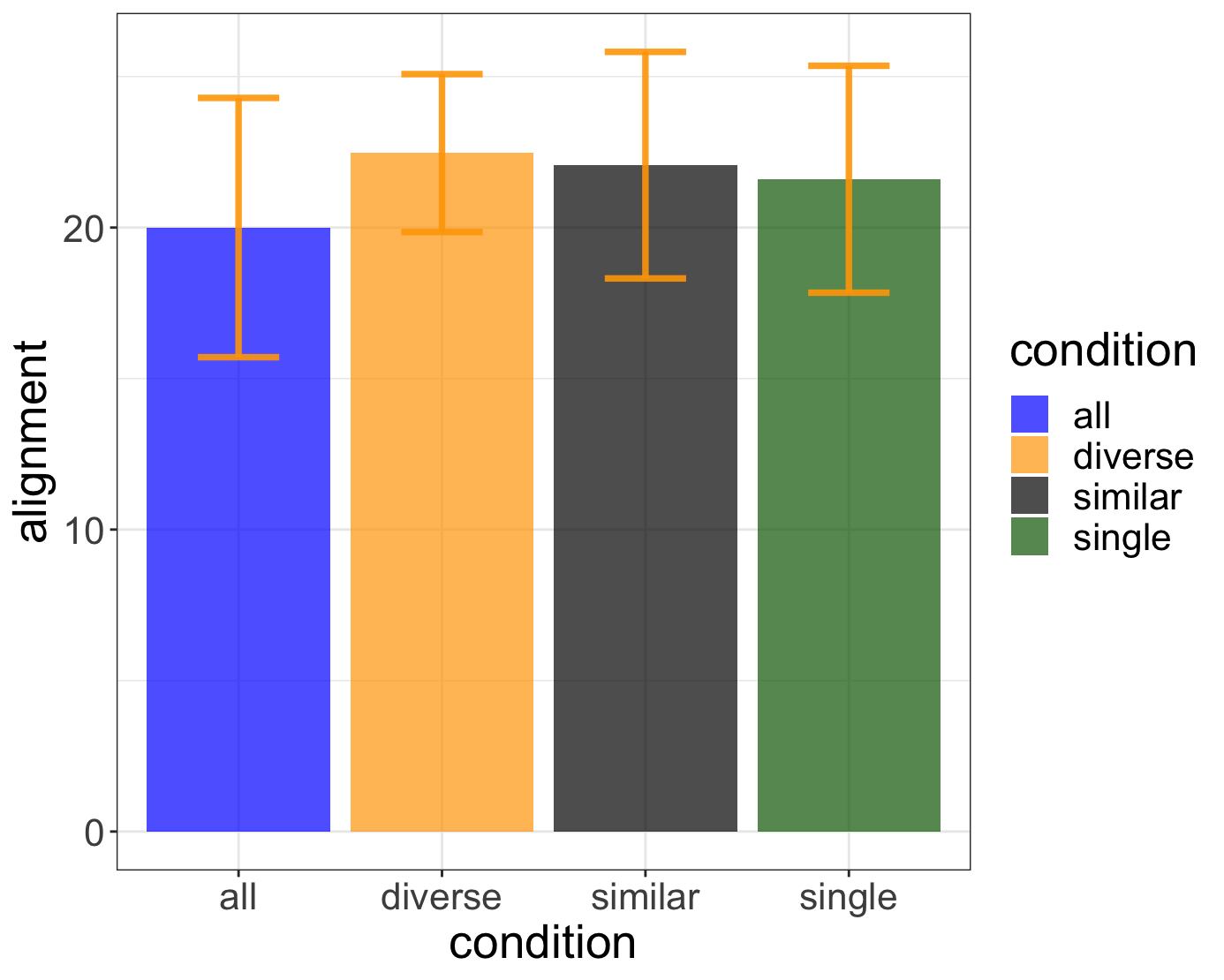}
         \caption{}
         \label{fig:alignment}
     \end{subfigure}
    \caption{Two bar graphs which show the performance of each study condition based on (a) usability and (b) alignment.}
     
    \label{fig:bar graphs}
\end{figure}

\begin{figure}[t]
  \centering
  \includegraphics[width=0.9\linewidth]{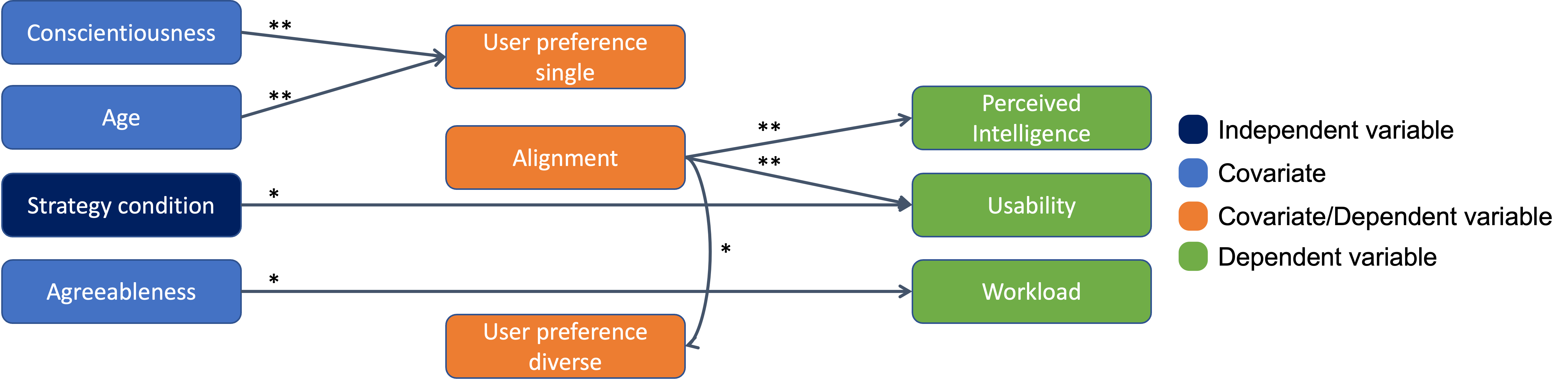}
  \caption{Summary of all the significant results in our study.}
  \label{fig:significance_plot}
\end{figure}

\subsubsection{RQ2}
\label{sec:Q1}
We first sought to measure the effect of the study condition, i.e. the method of recommending strategies to participants, on the usability, perceived intelligence, and workload. 
An ANOVA on our linear regression model for usability yielded a significant difference in usability across the recommendation types (F(3, 56) = 4.1609, p $<$ 0.05).
A Tukey post-hoc test showed that the single strategy recommendation was perceived to be significantly more usable than similar strategy recommendations (p $<$ 0.01). 
The single strategy was rated higher than the other two conditions as well (diverse, all), however the pairwise difference was not found to be significant. 
The method of recommending strategies was not found to be significant with respect to perceived intelligence or workload.


\subsubsection{RQ3}
Next, we sought to understand how intrinsic personality types affected preference towards any given strategy. 
Firstly, conscientiousness significantly impacted the usability of strategy recommendation systems in general.
Participants who had higher degrees conscientiousness (F(1,58) = 9.6539, p $<$ 0.01) tended to significantly perceive the recommendation system as more usable. 
A participant who likes to be more prepared and are attentive to smaller details may find it more usable to be recommended strategies for solving a complicated task because they are willing to spend the time to comprehensively assess their options.
Workload was significantly higher for participants who had higher traits of agreeableness (F(1, 58) = 4.3898, p $<$ 0.05). This is a logical finding, as agreeable participants would be more likely to simulate every single strategy to best understand each recommendation, thereby incurring a higher workload. 
Additionally, we found that age is positively correlated to preference towards single strategy recommendation systems (F(1, 58) = 14.6446, p $<$ 0.01). This could imply that older people are confident in their strategies and do not want to waste their time analyzing other options.

\begin{figure}[t]
     \centering
     \begin{subfigure}[b]{0.48\textwidth}
         \centering
         \includegraphics[width=\textwidth]{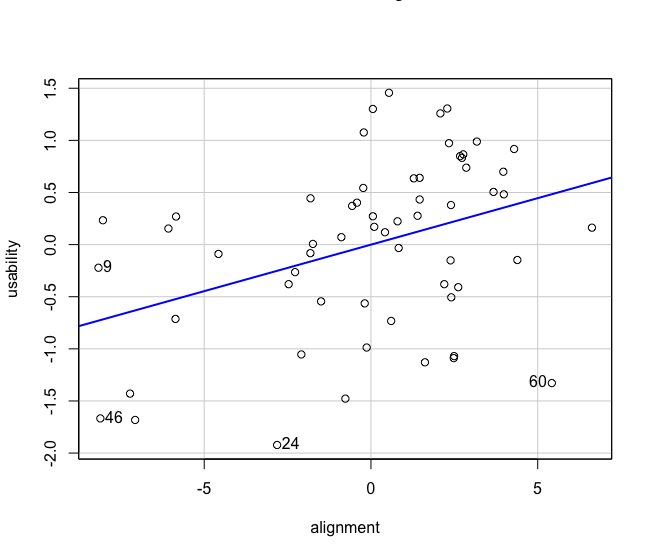}
         \caption{}
         \label{fig:cond_consc_upc2}
     \end{subfigure}
     \begin{subfigure}[b]{0.48\textwidth}
         \centering
         \includegraphics[width=\textwidth]{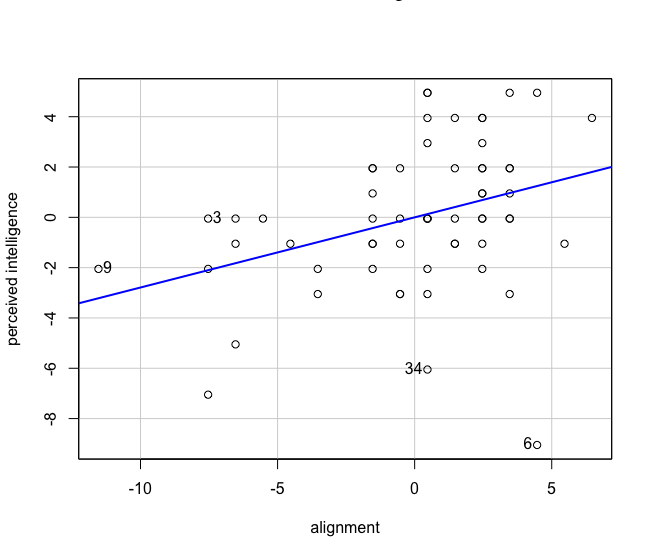}
         \caption{}
         \label{fig:cond_consc_upc4}
     \end{subfigure}
    \caption{These plots denote the impact of perceived alignment on usability and perceived intelligence.}
    \label{fig:bar graphs}
\end{figure}

\subsubsection{RQ4}
Next, we report some findings pertaining to the perceived alignment of recommendations with respect to their innate preferences. 
Alignment was found to significantly increase usability (F(1, 58) = 8.6770, p $<$ 0.01).  
From Section~\ref{sec:Q1}, we know that the single strategy condition was perceived as the most usable.
This finding is in line with prior work in personalization ~\citep{tan2008learning,chen2006recommendation, linden2003amazon} which suggests that the people generally like their preferences being reflected in the recommended options. 
Alignment also significantly impacted perceived intelligence (F(1,58) = 9.3313, p $<$ 0.01). 
We found that participants who felt that the recommendations provided were more aligned with their inherent preferences, perceived the system are more intelligent. 
It is logical that participants who felt that the system was able to accurately understand and encode their gameplay preferences into the recommendation ascribed more intelligence to the system.


Additionally, while modeling general preference of diverse strategies, through our preference questionnaire, alignment was a significant covariate (F(1, 58) = 5.0068, p $<$ 0.05). 
Higher perceived alignment improved a participants general preference towards diverse strategies, which indicates that when a participant received a recommendation which was perceived to more accurately reflect their preferences, they were more open to being provided other contrasting strategies. 

\subsection{Discussion}
Our experiment highlights key trends with regards to the design of strategy recommendation systems. Supporting prior work on general recommendation systems~\citep{nguyen2018user}, our results highlight the importance of accounting for personality factors during the design of AI strategy recommendations. We found that conscientiousness significantly reduced a participants preference towards the single strategy mode of presentation and agreeableness can increase the workload experienced. 
This finding hints at the possibility that the intrinsic traits of the end-user can be used to preemptively design more suitable strategy recommendation systems. For example, prior work has shown that high conscientiousness is very important for high-stress, skill based positions like surgeons or helicopter pilots~\citep{dickens2013looking, grice2006personality, mullola2018personality}.
Our results may indicate that humans in such occupations may not prefer the single strategy condition of recommendations.
Also, physicians working in the private sector with general practice or occupation health as their specialization, can have high levels of agreeableness~\citep{mullola2018personality}. Professionals in these fields might experience higher workload while using a strategy recommendation systems.
Additionally, there seemed to be a general preference towards single strategy recommendation systems over similar strategies recommendation system. Thus, if a recommendation system is capable of inferring user preferences, users may find it more usable to be presented with just the most relevant strategy instead of choosing from a list of strategies similar to the most relevant strategy.


Prior work has shown that humans hold AI-assistants to a higher standard than human-assistants~\citep{chen2021err}. To design recommendation systems for human-AI collaborative tasks, that humans will be willing to adopt, we need to better understand the perception of these systems. 
Our results provide salient insights regarding the usability and perceived intelligence of strategy recommendation systems.
The participants who feel that their preferences are aligned with their recommendations perceive the system as more intelligent and usable. 
In this work, we provide a proof of concept of the benefits of aligning recommendation options with user preferences, future work should explore automatically inferring user's innate preferences at scale and aligning strategy recommendations with it. 
Additionally, our qualitative questionnaire suggests incorporating explainable AI mechanisms \citep{zhu2020effects, ehsan2019automated, silva2022explainable} in strategy recommendation systems could further benefit confidence in the system.
For instance, one participant commented that, ``I wish that the interface would explain a bit more on how my choices were converted into the strategy so I would feel more confident in it.''
We also provide insights regarding the impact of the study condition on each of these dependent variables to offer more insight into modelling these dependent variables. An important next step would be to study these measures from the perspective of user-adoption, in order to ascertain how each of these three properties (alignment, usability, and perceived intelligence) affect willingness to adopt a given mode of recommending strategies. 

\section{Limitations}
Firstly, although we provide participants with a tutorial as well as a simulation of Risk gameplay, we cannot be certain that participants have adequately understood Risk. We obtain verbal confirmation from participants that they are comfortable with the game, however, we do not specifically test for this. 
Without adequately understanding how to play Risk, participants may not be able to create a mental model of the strategy they would employ, and thus may have difficulty evaluating the strategy recommendations. 
While important to acknowledge, this does not diminish our findings and analysis due to the number of participants who took part in our experiments and the comprehensive nature of the tutorial and simulation. 

Secondly, in this study we have not conducted factor analysis of the preference questionnaire we designed. Our analysis showed significant trends relating to the participant preference, that provided support for the validity of our questionnaire. 
However, in future work, we hope to perform factor analysis to ensure reliability of the questions in relation to each factor (Single, Similar, Diverse, All). 
Lastly, while our study provides novel insights regarding broad strategy recommendation preferences, our experiment is not setup in an immersive real-world task.
In future work, we hope to develop a real-world collaborative experiment with a robot wherein the robot recommends strategies to a participant, and the human employs the recommendations to specify how the robot should complete the collaborative task. We would also conduct this human-subjects experiment on a more diverse population in the future.

\section{Conclusion}
As humans collaborate with AI-agents to solve more challenging tasks, humans may not have the capability to translate their preferences into actionable strategies that an AI agent can execute. 
To solve this issue, we need to develop strategy recommendation systems that can take in a user's preferences and recommend well defined strategies. 
Prior work on user-centric recommendation systems has studied the impact of personality type of a user on the perception of a recommendation system as well as the general structure of a recommendation list. However, if the task at hand is more intricate and complex these preferences might change. In this paper, we conducted a novel human subjects experiment to understand how differing types of recommendations impact the usability, perceived intelligence, workload and preference towards any given mode of recommendation. 
We created and validated a novel method of encoding a participant's preferences into an actionable strategy for Risk through three simple gameplay questions.
We also developed a new questionnaire to gauge user preference towards receiving \textit{single}, \textit{similar}, \textit{diverse}, or \textit{all} types of recommendations. Our analysis showed that certain personality traits (e.g.: conscientiousness) have a significant impact on the preference of a particular type of recommendation system. 
Finally, we report a correlation between preference, alignment and usability, highlighting the need for further mechanisms to automatically infer user preferences and align strategy recommendations with it.
Our work provides insights into user preferences for a strategy recommendation system which can be used to design personalized systems for human-AI teaming in the future. 
Personalization can in turn lead to higher user satisfaction and adoption of the strategy recommendation system.


\section{Acknowledgement}
This work was supported by the Office of Naval Research under grant N00014-19-2076 and a gift by Konica Minolta to the Georgia Tech Research Foundation.



\bibliographystyle{ACM-Reference-Format}  
\bibliography{references}  

\appendix
\section{Additional Results}
We utilized AICc as our Occam's razor to finalize the model for analyzing each dependent variable. 
For each dependent variable, we began with a model that modelled each individual covariate and independent variable along with all pairwise interaction effects between these variables. 
We removed variables until we arrived at the linear regression model with the lowest AIC score. 
In this section, we report an additional significant finding that was not relevant to our research questions.

\begin{figure}[t]
     \centering
     \begin{subfigure}[b]{0.48\textwidth}
         \centering
         \includegraphics[width=\textwidth]{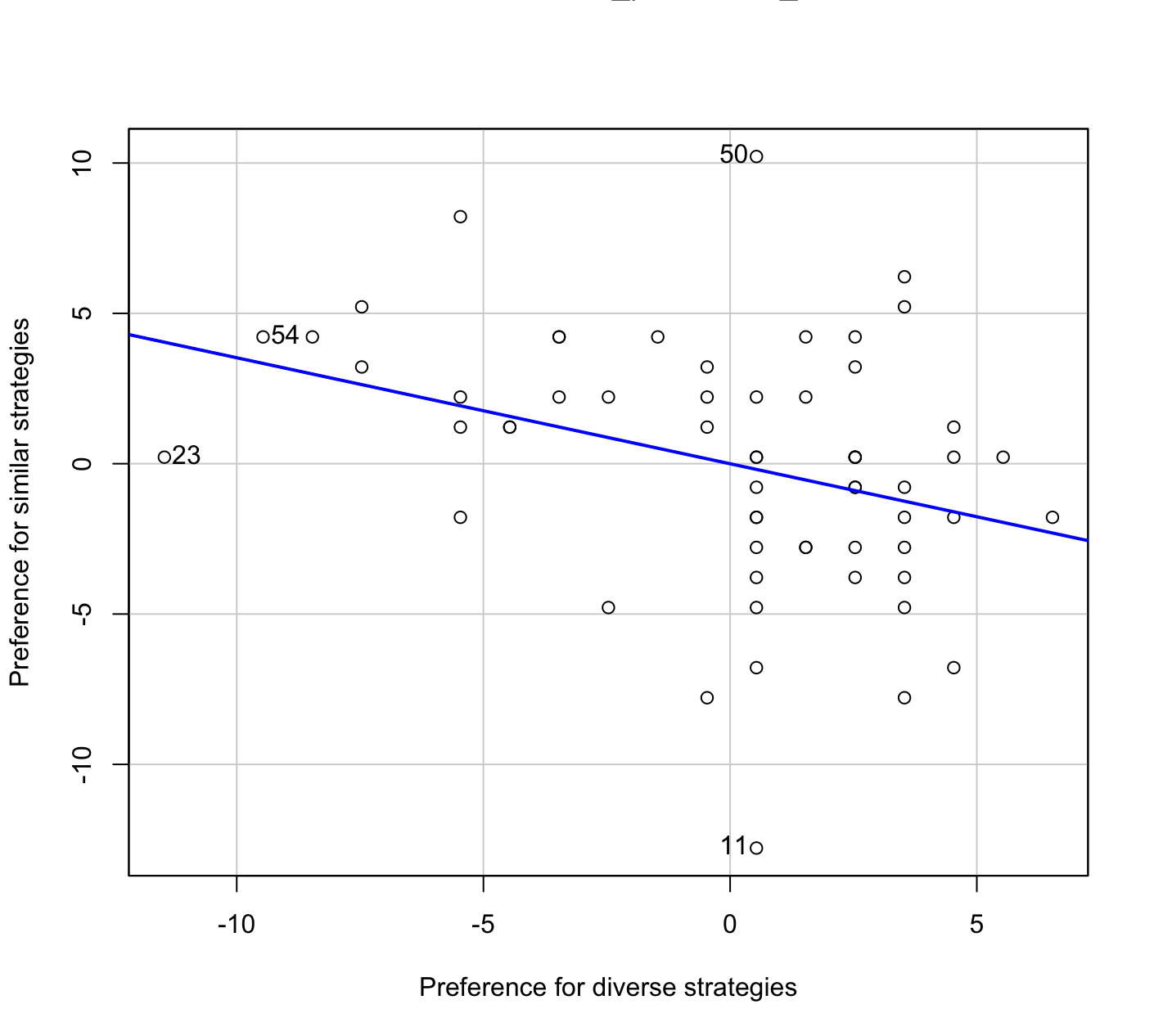}
         \caption{}
         \label{fig:simvdiv}
     \end{subfigure}
     \begin{subfigure}[b]{0.48\textwidth}
         \centering
         \includegraphics[width=\textwidth]{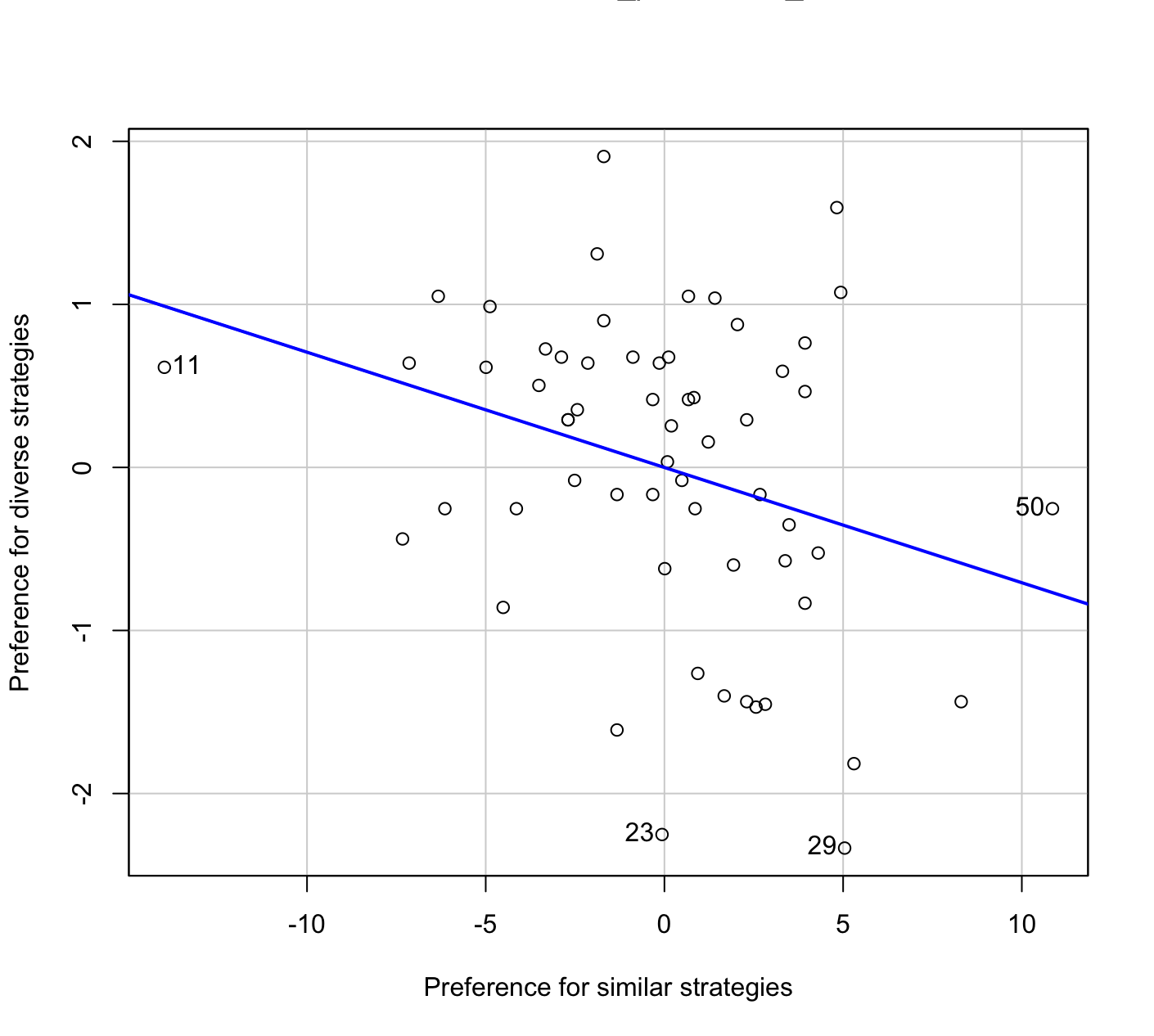}
         \caption{}
         \label{fig:divvsim}
     \end{subfigure}
    \caption{These plots denote the correlation between preference for diverse strategies and preference for similar strategies.}
    \label{fig:add_results}
\end{figure}

We found that the preference for diverse strategies was inversely related to the preference for similar strategies. This was found to be a significant correlation while modeling both preference for similar strategies (F(1, 58) = 7.9319, p $<$ 0.01) and preference was diverse strategies (F(1, 58) = 7.6388, p $<$ 0.01) (Figure~\ref{fig:add_results}).
This result is intuitive as people who like to recommended strategies aligned to their preference would not like to be recommended diverse strategies and vice-versa.

\section{Questionnaires}
\subsection{Calibration Questionnaire}
In this section, we provide the seven questions utilized to encode the preferred gameplay style of a participant. 
The calibrated gameplay style was further used to recommend strategies (Figure~\ref{fig:rec_example}). The questionnaire comprises of seven questions each containing two options to choose from (Figure~\ref{fig:calibration}). 

\subsection{Preference Questionnaire}
This section contains the details regarding the questions and scoring of our novel preference questionnaire. We have four questions directed at measuring the preference towards each type of condition. We randomized the questions for a particular condition.
\paragraph{\textbf{Questionnaire}}
Imagine you are completing a challenging task and you have to come up with a well-defined strategy that you can use. You are then provided with an AI agent which can understand your preferences and accurately provide strategy recommendations (similar to the process you went through in this study). In such instances, please answer the following questions regarding how you would like to receive plan recommendations on a scale of 1-7, 1 being strongly disagree and 7 being strongly agree.

\begin{enumerate}
    \item I would like to be presented with a single plan which best reflects my preferences. 
    \item Even if I am presented with a plan that best represents my preferences, I would prefer receiving additional options that I could consider.
    \item I don’t need more than one plan if I am presented with a plan which best reflects my preferences.
    \item Evaluating more than one plan is not worth the extra work.
    \item I would like to be presented with multiple plans: one that best reflects my strategy preferences and a couple of alternative plans that are slightly different.
    \item I would not like to be presented with more than one strategy related to my preferences. The other options should be dissimilar.
    \item I would not want to be presented with a diverse set of plans. Instead, I would want a few plans that are closely aligned with my strategy preferences.
    \item Picking from a set of similar plans is better than picking from a set of dissimilar plans.
    \item I would like to be presented with multiple, diverse plans: one that best reflects my strategy preferences, one that is very different from my preferences, and one that is neither similar nor dissimilar from my preference.
    \item I would prefer for all the plans presented to me to be similar rather than dissimilar.
    \item I want to be able to consider a diverse set of options if the list is not too long.
    \item I would not like to consider plans that are different from my preferences.
    \item I would like to be presented with a list of every possible plan – one for each possible strategy.
    \item Evaluating all possible plans is not worth the effort.
    \item Seeing the breadth of planning possibilities is helpful for identifying the best plan.
    \item Being presented with all possible plans will not help me in accomplishing my goals.
\end{enumerate}

\paragraph{\textbf{Scoring}}
This section covers how we computed scores for each factor in the questionnaire. Some items in the questionnaire need to be reversed prior to scoring. Each factor in our questionnaire had at least one reverse item to ensure that participants were paying attention to the questions. 
Items 2, 6, 10, 12, 14, and 16 need to be reversed before scoring:\newline
$
1 = 7\\
2 = 6\\
3 = 5\\
4 = 4\\
5 = 3\\
6 = 2\\
7 = 1\\$
Preference for Single Strategy Recommendation : Sum of items 1, 2r, 3, 4\\
Preference for Similar Strategies Recommendations :  Sum of items 5, 6r, 7, 8\\
Preference for Diverse Strategies Recommendations :  Sum of items 9, 10r, 11, 12r\\
Preference for All Strategies Recommendation :  Sum of items 13, 14r, 15, 16r\\
r refers to a reverse scaled item here  

\section{Interview Questions for Calibration Study}
During the calibration phase if the participant preferred the reverse strategy (opposite of the strategy generated by our decision tree), we asked them these semi-structured interview questions to understand the missing gap. 
\begin{enumerate}
    \item What strategy do you generally follow in the game?
    \item Which strategy is most aligned with your strategy? 
    \item How do you think Strategy A is different from Strategy B?
    \item Can you rank the following strategies from least aligned to most aligned with your strategy? 
    \item Why did you select option X?
    \item What do you think this question is trying to convey?
    \item What changes in the setup would make you change your answer? 
\end{enumerate}

\section{Data Filtering Rubric}
This section contains the rubric that was used to filter out the data for our main study. We discarded four data points using the following rubric - 
\begin{enumerate}
    \item The participant did not read through the goals and constraints of the recommended strategy. 
    \item The participant evaluated the Qualtrics UI rather than the recommendation system. 
    \item Instead of evaluating the strategy the participant only evaluated how well the simulated strategy was performing.
    \item The participants made some mistakes while going through the user study.
\begin{enumerate}
    \item For e.g. - The participant selected a wrong options in the initial calibration questionnaire
    \item For e.g. - The participant forgot to answer one of the likert items of the preference questionnaire
\end{enumerate}

\end{enumerate}

\begin{table}
\centering
  \caption{This table details the independent variable, dependent variable and covariates for each model. We have also listed down the assumptions of the ANOVA test and transforms applied.}
  \vspace{15px}
  \label{tab:stat_tests}
  \begin{tabular}{|p{0.2\linewidth} | p{0.15\linewidth} | p{0.3\linewidth}| p{0.15\linewidth}| p{0.10\linewidth}|} 
    \hline 
    DV & Transform & IV/Covariates & Shapiro-Wilk & Levene's \\ \hline
     Usability& boxcox & Condition, Conscientiousness, Openness, Preference for Single Strategy, Alignment & $p=0.07245$ & $p=0.3248$ \\\hline 
     Alignment& boxcox & Preference for Diverse Strategies & $p= 0.1011$ & $p=0.3682$ \\\hline 
     Preference for Single Strategy& N/A & Conscientiousness, Age    & $p= 0.6839$ & $p=0.8932$ \\ \hline 
     Preference for Similar Strategies& N/A & Preference for Diverse Strategies & $p= 0.3213$ & $p=0.3786$ \\ \hline 
     Preference for Diverse Strategies& boxcox & Preference for Similar Strategies, Alignment & $p= 0.4235$ & $p=0.4235$ \\ \hline 
     Workload& N/A & Agreeableness, Conscientiousness, Preference for Single Strategy, Age& $p= 0.5515$ & $p=0.5909$ \\ \hline 
     Perceived Intelligence& boxcox & Alignment & $p=  0.4376$ & $p=0.8405$ \\ \hline 
     
    \end{tabular}
\end{table}

\begin{figure}[!b]%
\centering
\subfloat[\centering Question 1]{{\includegraphics[width=0.4\linewidth]{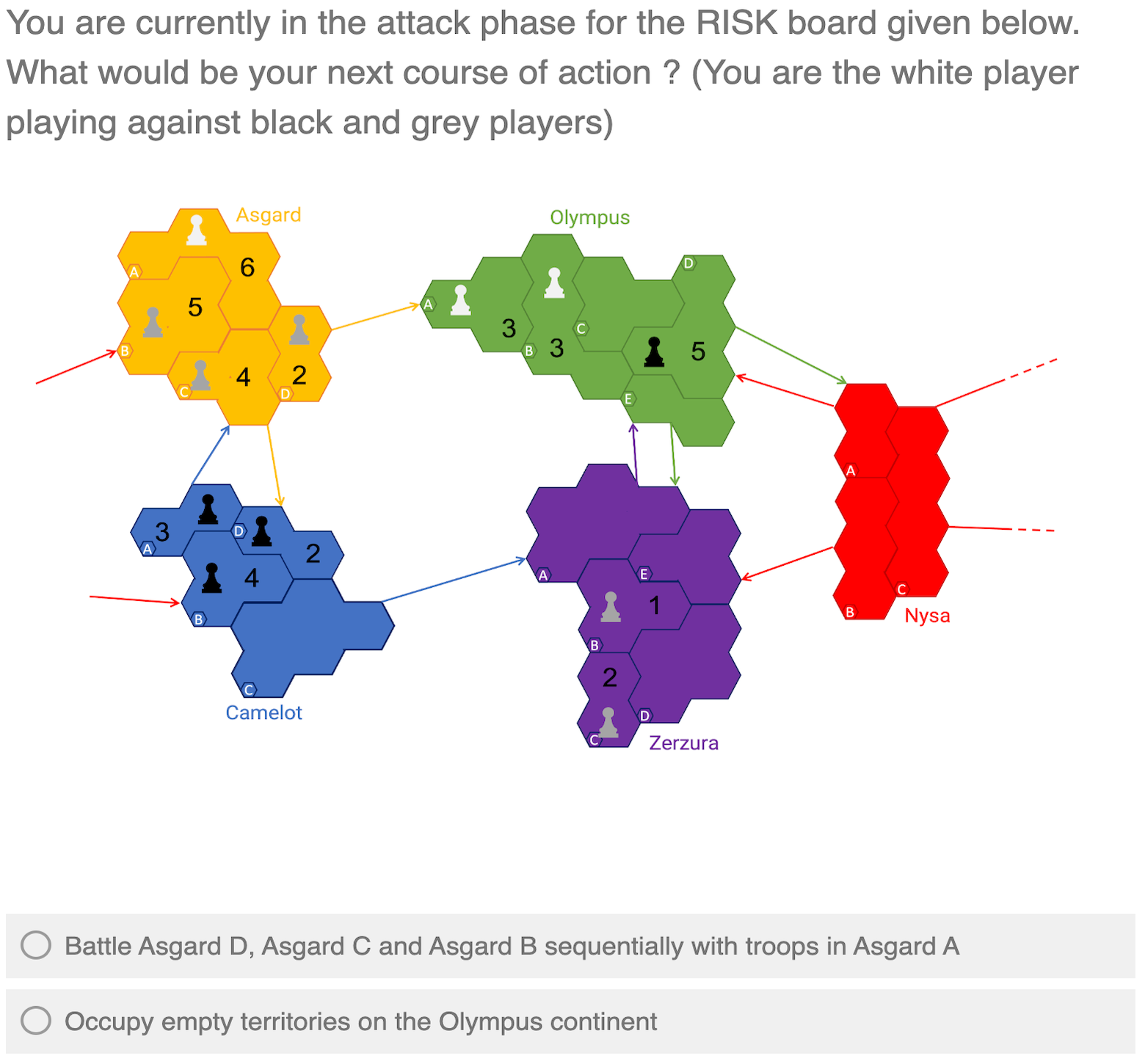} }}%
\qquad
\subfloat[\centering Question 2]{{\includegraphics[width=0.4\linewidth]{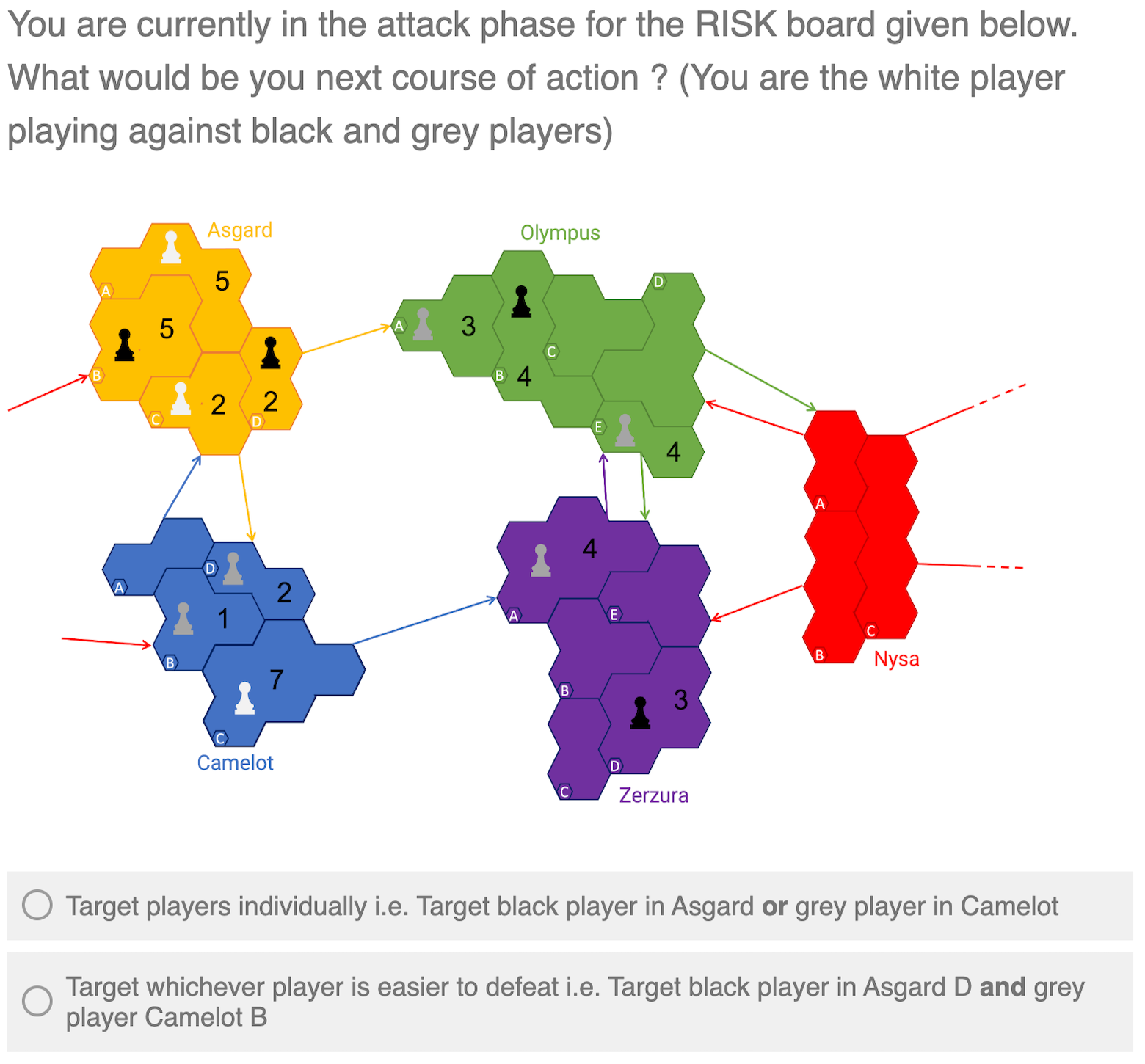} }}\
\\
\subfloat[\centering Question 4]{{\includegraphics[width=0.4\linewidth]{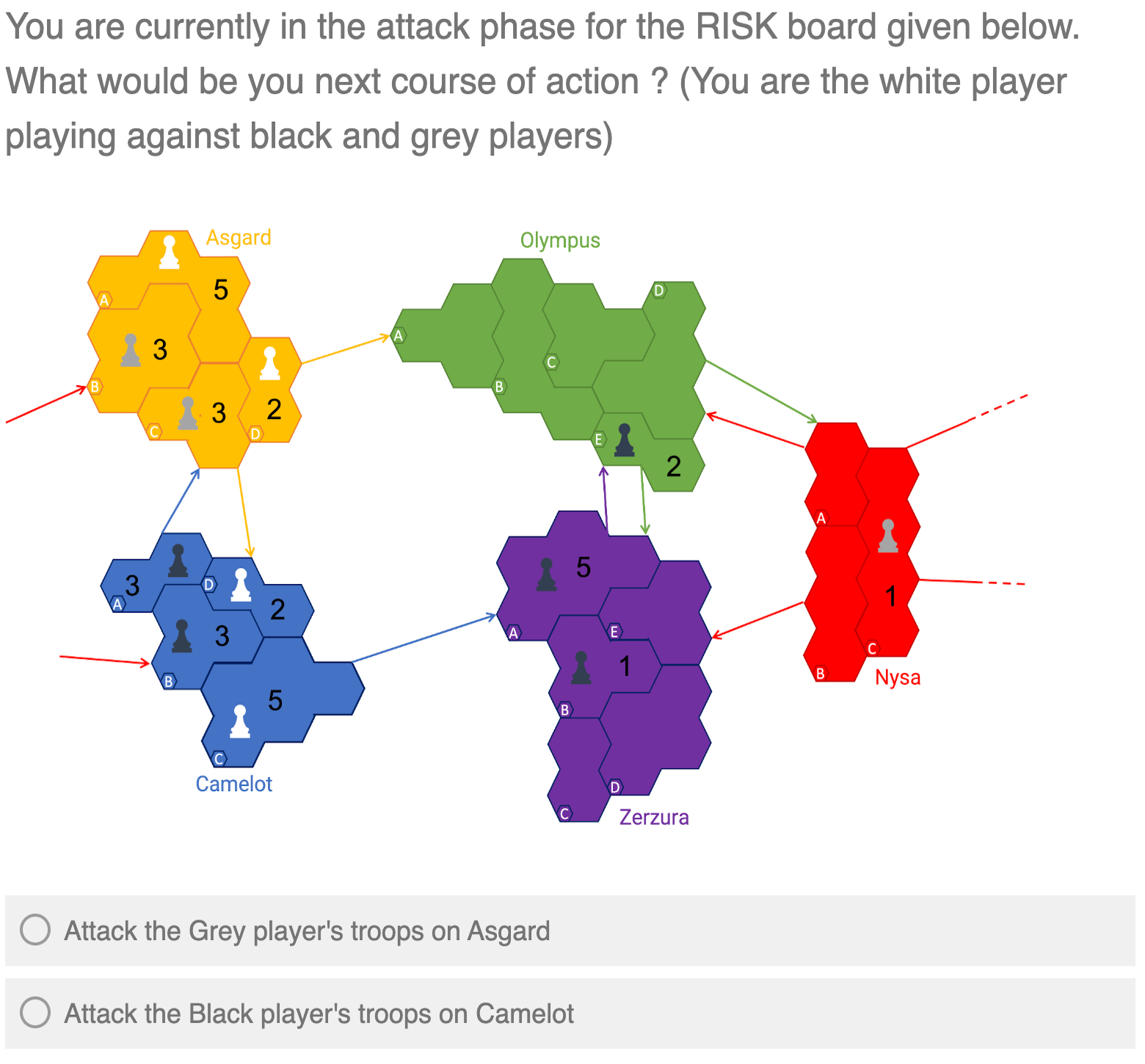} }}%
\qquad
\subfloat[\centering Question 6]{{\includegraphics[width=0.4\linewidth]{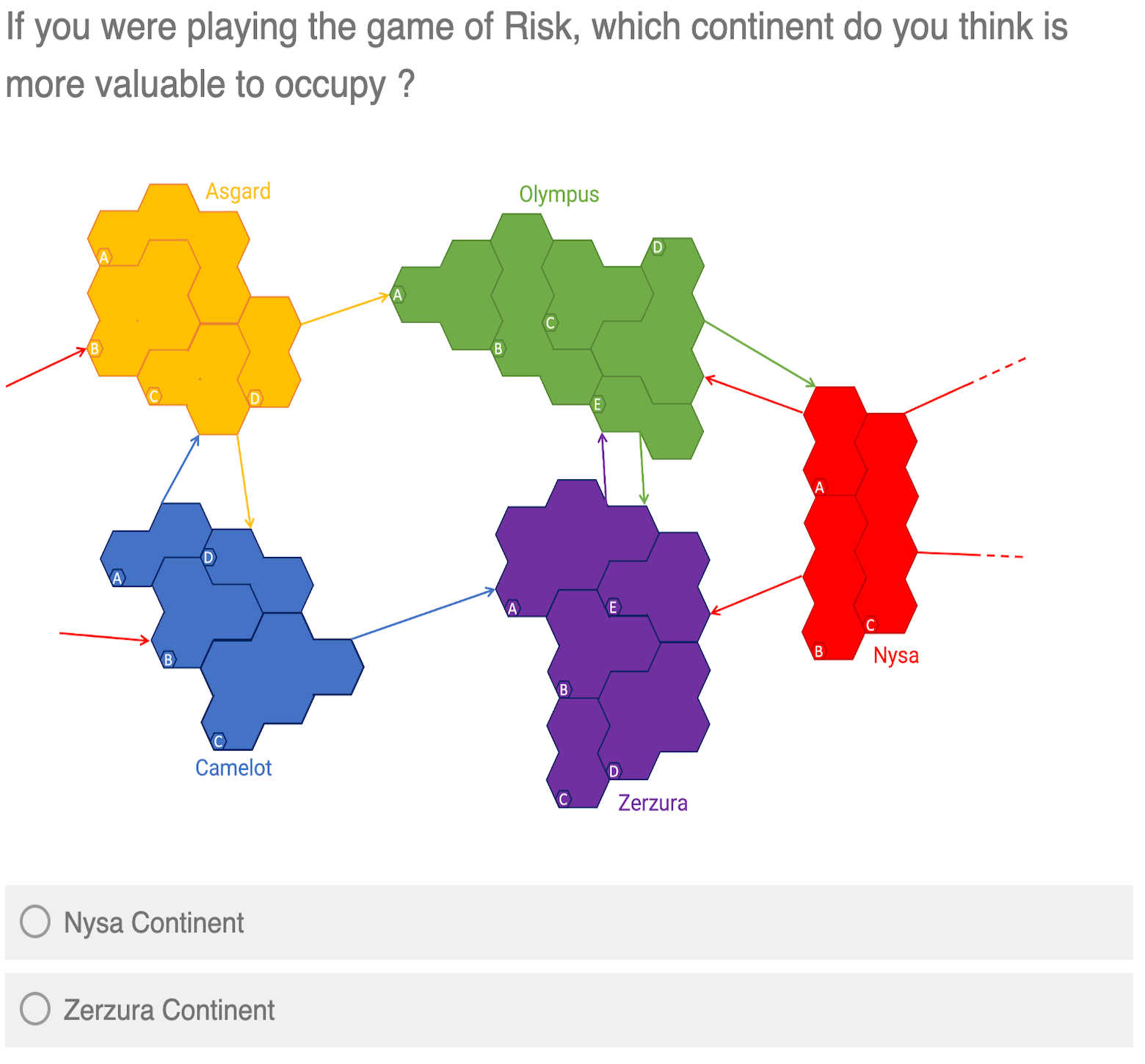} }}\
\end{figure}%
\begin{figure}\ContinuedFloat%
\centering
\subfloat[\centering Question 3]{{\includegraphics[width=0.4\linewidth]{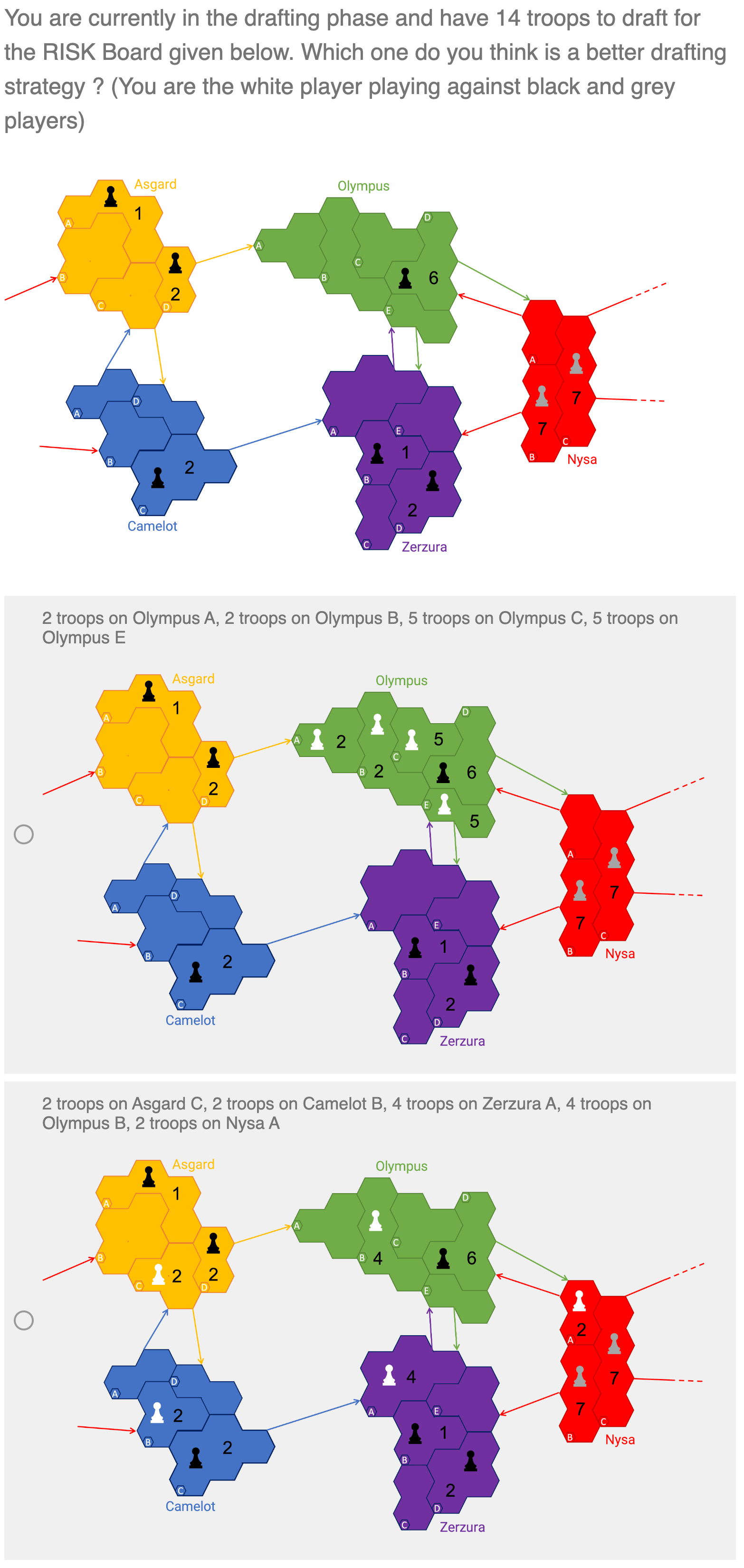} }}%
\qquad
\subfloat[\centering Question 5]{{\includegraphics[width=0.4\linewidth]{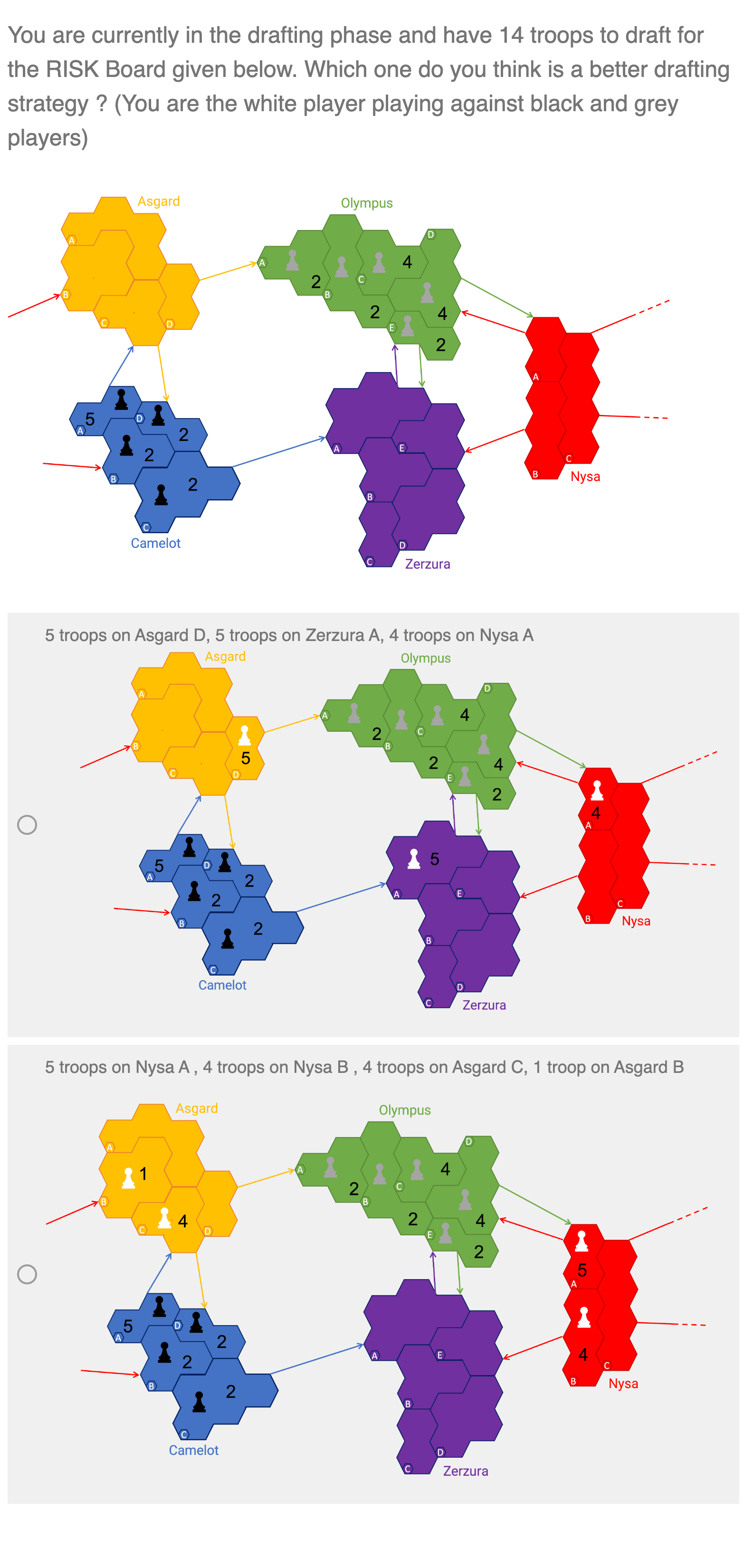} }}%
\end{figure}
\begin{figure}\ContinuedFloat%
  \centering
  \subfloat[\centering Question 7]{{\includegraphics[width=0.4\linewidth]{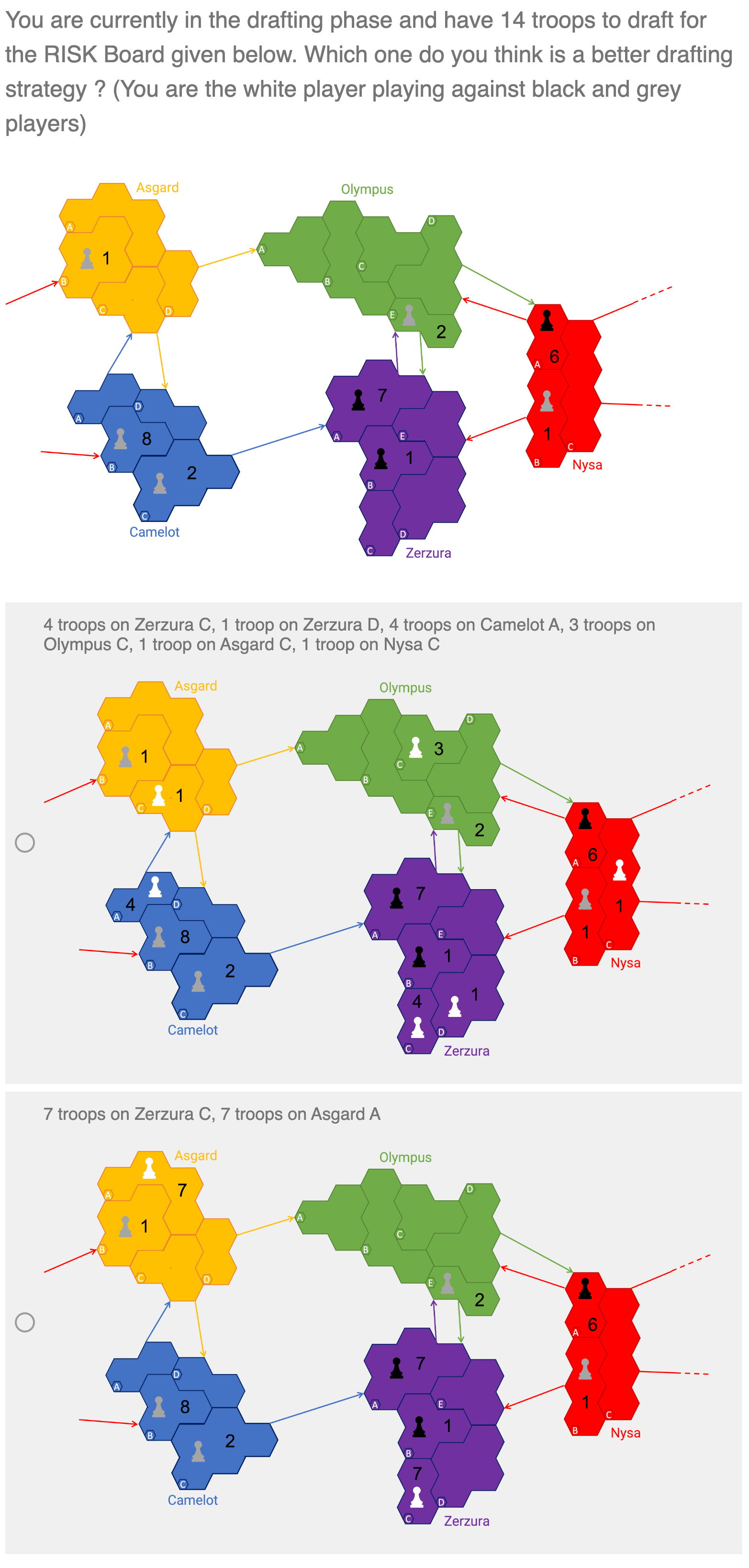} }}\
  \caption{Calibration Questionnaire}%
  \label{fig:calibration}
\end{figure}






\end{document}